\documentclass[journal]{IEEEtran}

\usepackage{xcolor,soul,framed} %,caption
\usepackage[pdftex]{graphicx}
\usepackage{amsmath,amssymb,amsfonts,amstext,bbm}
\usepackage{array}

\usepackage{flushend}

\usepackage{multicol}
\usepackage{caption}
\usepackage{subcaption}
\usepackage{array}

\usepackage{tikz}  						
\usetikzlibrary{shapes.geometric, arrows,calc}
\usetikzlibrary{positioning}
\usetikzlibrary{decorations.pathreplacing}
\tikzstyle{arrow} =[thick,->,>=stealth] % thick lines and stealth

\usepackage{pgfplots}
\pgfplotsset{compat=newest}
\usetikzlibrary{plotmarks}
\usetikzlibrary{arrows.meta}
\usepgfplotslibrary{patchplots}
\usepackage{grffile}
\usepackage{amsmath}
\usepackage{graphicx}

\newcommand{\Exp}{\mathbb{E}} 

\newcommand{\TExp}[1]{\mu_{\scriptscriptstyle #1}}
\newcommand{\TVar}[1]{\sigma^2_{\scriptscriptstyle #1}}

\newcommand{\WE}{G_i^\lambda}
\def\define{\stackrel{\Delta}{=}}

\begin{document}
\bstctlcite{IEEEexample:BSTcontrol}
    \title{Exponentially-Weighted Energy Dispersion Index for the Nonlinear Interference Analysis of Finite-Blocklength Shaping}

\author{\IEEEauthorblockN{Kaiquan~Wu,~\IEEEmembership{Student Member,~IEEE}
Gabriele~Liga,~\IEEEmembership{Member,~IEEE}
Yunus~Can~G\"ultekin~\IEEEmembership{Member,~IEEE}, and 
Alex~Alvarado~\IEEEmembership{Member,~IEEE}}\\
\IEEEauthorblockA{Eindhoven University of Technology, 5600 MB Eindhoven, The Netherlands}% <-this % stops an unwanted space
\thanks{Corresponding author: K. Wu (email: k.wu@tue.nl).}}

% ====================================================================
\maketitle

% === ABSTRACT ====================================================================
% =================================================================================
\begin{abstract}
A metric called exponentially-weighted energy dispersion index (EEDI) is proposed to explain the blocklength-dependent effective signal-to-noise ratio (SNR) in probabilistically shaped fiber-optic systems. EEDI is better than energy dispersion index (EDI) at capturing the dependency of the effective SNR on the blocklength for long-distance transmission.
\end{abstract}
\IEEEpeerreviewmaketitle

\section{Introduction}
Probabilistic amplitude shaping (PAS) \cite{bocherer2015bandwidth} can realize near capacity-achieving transmission for the additive white Gaussian noise (AWGN) channel. In fiber optical communications, significant shaping gains over uniform quadrature amplitude modulation (QAM) are achieved by the AWGN-optimal PAS \cite{buchali2016rate,fehenberger2016probabilistic}. However, these gains are undermined by the nonlinear interference (NLI) penalty, since shaping can enhance NLI effects with respect to uniform signaling \cite{renner2017experimental}. This penalty is enhanced by i) using a Maxwell-Boltzmann distribution, and ii) the \emph{temporal} structure of the transmitted shaped symbols. The penalty from the former can be reduced by optimizing the distribution of the constellation symbols to be more NLI-tolerant \cite{renner2017experimental,cho2016low,sillekens2018simple}.~The penalty from the latter can be instead mitigated by manipulating the temporal structure of the symbol sequences \cite{dar2014shaping,yankov2017temporal}.~The amplitude shaper in PAS imposes a hidden temporal structure on the symbols, and thus, the symbols can no longer be treated as independent identically distributed (i.i.d.).~It was found that the temporal structure caused by short shaping blocklengths can provide effective signal-to-noise ratio (SNR) gains, due to a weaker presence of nonlinearities \cite{amari2019introducing}. Therefore, a straightforward approach would be to simply employ short blocklengths \cite{geller2016shaping,fehenberger2020mitigating}. The NLI mitigation is intuitively explained by the fact that using short blocklengths avoids multiple consecutive occurrences of high-energy symbols, and thus, induces less NLI \cite{fehenberger2019analysis,skvortcov2020huffman}. 

Recently, in \cite{kaiquan2021EDI} we analyzed the statistical properties of symbols generated by constant composition distribution matching (CCDM) \cite{schulte2015constant} with finite blocklengths. Inspired by the behavior of time-domain first-order perturbation NLI models \cite{mecozzi2012nonlinear} and the finite memory Gaussian noise (GN) model \cite{agrell2014capacity}, we found that the variations of the windowed symbol energy are crucial for the NLI generation. We also proposed in \cite{kaiquan2021EDI} a precise metric to quantify the effect of energy variations on the NLI magnitude, the so-called \textit{energy dispersion index} (EDI). One drawback of EDI is that all symbol energies within a time window are assumed to be equally important. This assumption does not reflect the reality as interfering symbols far away from the symbol of interest are expected to have smaller impact on the NLI impinging on that symbol than those nearby. %This uneven contribution has been captured by the perturbation terms in the first-order perturbation model.

In this paper, we propose a refined version of the EDI, which we call exponentially-weighted EDI or EEDI. EEDI takes into account the fact that the NLI contribution from different symbol energies varies depending on their relative delay with respect to the symbol of interest. Our contribution in this paper is to verify that by weighting the interfering symbol energies properly, EEDI is a better effective SNR estimator than EDI at long transmission distances.

\section{Weighted Energy in Fiber Channel Model}

In our previous study, we assumed that the NLI generation is dominated by symbol energies within a finite time window \cite{kaiquan2021EDI}.~In this paper, we extend this time window to infinity and we introduce a decay factor to weigh the effect of temporal separation between symbols on the NLI generation. Using the first-order perturbation model and assuming single channel transmission, the NLI term $Z_{\text{NLI},0}$, which is modeled as additive noise on the transmitted symbol $X_0$, can be expressed as \cite[Eq.~(57)]{mecozzi2012nonlinear}

\begin{equation}\label{KnMd0}
   Z_{\text{NLI},0}= \jmath \gamma \sum_{h=-\infty}^{\infty} \sum_{j=-\infty}^{\infty} \sum_{l=-\infty}^{\infty} \! S_{h, j, l} X_{h} X_{j} X_{l}^{*}.
\end{equation}

In (\ref{KnMd0}), $\gamma$ is the nonlinear coefficient, and the complex kernels $S_{h, j, l}$ determine the self-phase modulation contribution of the symbol triple product $X_{h} X_{j} X_{l}^{*}$ based on the temporal separation of its factors.~Note that the NLI term for the cross-channel interference can be expressed in the same form as in (\ref{KnMd0}) \cite[Eq.~(60)]{mecozzi2012nonlinear}. Since the perturbation terms $S_{h, j, l}$ satisfying $j=l$ have large magnitude \cite[Sec.~II-B]{dar2014shaping}, we can approximate (\ref{KnMd0}) as

\begin{equation}\label{pmd}
  Z_{\text{NLI},0}\approx \jmath \gamma \, \sum_{h=-\infty}^{\infty} X_{h} \underbrace{\sum_{l=-\infty}^{\infty} S_{h, l, l} |X_{l}|^2.}_\text{Weighted Sum of Energies}
\end{equation}

It can be seen in (\ref{pmd}) that the NLI experienced by $X_{0}$ is determined by the weighted sum of symbol energies.~The variance of this term determines the variance of the induced NLI. This term indicates that all transmitted symbols generate NLI proportional to their energies.
However, these energies are weighted by $S_{h, l, l}$, whose magnitude change as a function of the index $l$ \cite{tao2011multiplier}. In general, $S_{h, l, l}$ slowly decays as the offset $|l|$ increases \cite{dar2014shaping}.

\section{Exponentially-Weighted EDI}
\begin{figure}[t]
  \centering
        \resizebox{1\linewidth}{!}{% % \documentclass[tikz,border=5pt]{standalone}
% \documentclass[tikz,float=false,crop=true]{standalone}
% \usepackage{pgfplots}
% \usepackage{amsmath, amssymb, amscd, amsthm, amsfonts}
% \usetikzlibrary {arrows.meta,shapes.multipart}
% \usetikzlibrary{shapes,arrows,positioning,calc,decorations.pathreplacing}
% \pgfplotsset{compat=newest}

% \newcommand{\Exp}{\mathbb{E}} 
% \newcommand{\Energy}{E}
% \begin{document}

\tikzstyle{symBlk0} = [rectangle,draw=black, line width = 0.7pt, minimum height=25pt, minimum width=30pt,inner sep = 0pt]

\tikzstyle{symBlk} = [rectangle,draw=black, line width = 0.7pt, minimum height=25pt, minimum width=30pt,inner sep = 0.1pt]

\tikzstyle{cBlk} = [rectangle, minimum height=25pt, minimum width=1.3pt,inner sep = 0pt]

\begin{tikzpicture}

  \colorlet{color min rgb}[rgb]{red}
  \colorlet{color max rgb}[rgb]{lime}

  \def\min{0}
  \def\max{180}

\foreach \i in {1,2,...,135}{
    \pgfmathsetmacro\myvalue{\i}

    \pgfmathtruncatemacro\lambda{(\myvalue-\min)/(\max-\min)*100}
    \colorlet{my color rgb}[rgb]{color min rgb!\lambda!color max rgb}
    
    \node[cBlk,fill=my color rgb] () at (\i pt+15pt,0) {};
}

\foreach \i in {135,136,...,270}{
    \pgfmathsetmacro\myvalue{\i}

    \pgfmathtruncatemacro\lambda{(270-\myvalue-\min)/(\max-\min)*100}
    \colorlet{my color rgb}[rgb]{color min rgb!\lambda!color max rgb}
    
    \node[cBlk,fill=my color rgb] () at (\i pt+15pt,0) {};
}

\foreach \i in {1,2,...,9}{
    \node[symBlk0] (sym\i) at (30*\i pt,0) {};
}

%%%%% different notation

\node[symBlk] () at (sym1) [label=below:$...$] {$...$};
\node[symBlk] () at (sym2) [label=below:$\lambda^{|l|}$] {$X_{i-|l|}$};
\node[symBlk] () at (sym3) [label=below:$...$] {$...$};
\node[symBlk] () at (sym4) [label=below:$\lambda$] {$X_{i-1}$};
\node[symBlk] () at (sym5) [label=below:$1$] {$X_{i}$};
\node[symBlk] () at (sym6) [label=below:$\lambda$] {$X_{i+1}$};
\node[symBlk] () at (sym7) [label=below:$...$] {$...$};
\node[symBlk] () at (sym8) [label=below:$\lambda^{|l|}$] {$X_{i+|l|}$};
\node[symBlk] () at (sym9) [label=below:$...$] {$...$};

\draw [decorate, line width = 1 pt, decoration={brace,amplitude=10pt,raise=3pt}] (sym1.north west) -- (sym9.north east) node [midway,above=12pt,text width=180pt] {$G_i^\lambda=\ldots+ \lambda^{|l|} |X_{i-|l|}|^2+\ldots+\lambda |X_{i-1}|^2 + |X_i|^2+\lambda |X_{i+1}|^2+\ldots+ \lambda^{|l|} |X_{i+|l|}|^2+\ldots$};

\end{tikzpicture}

% \end{document}
}
    \caption{An illustration of the weighted sum of energy $G_i^\lambda$}
    \label{expWndEng}
\end{figure}

To reflect the effect of the $S_{h, l, l}$ in (\ref{pmd}), we heuristically assume that its magnitude decays exponentially with increasing $|l|$.~Let $\lambda$ be a forgetting factor, where $0\leq\lambda\leq 1$, and let $X_{l}$ be the symbol $|l|$ symbol periods away from $X_{0}$. We assume that the NLI contributions associated with $|X_{l}|^2$ can be expressed as $|S_{h, l, l}|=\lambda^{|l|} |S_{h, 0, 0}|$. 
Then, we design a variable $G_i^\lambda$ to capture the weighted sum of symbol energies around symbol $X_i$, which is defined as
\begin{equation}\label{weng}
  G_i^\lambda\define\sum_{l=-\infty}^{\infty} \lambda^{|l|} |X_{i+l}|^2. 
\end{equation}
We can then express the NLI $Z_{\text{NLI},i}$ in the product between $G_i^{\lambda}$ and $\sum_{h=-\infty}^{\infty} X_{h+i}|S_{h+i, 0, 0}|$ which is still expected to be somehow correlated to $G_i^{\lambda}$.

Fig.~\ref{expWndEng} illustrates how $G_i^\lambda$ is obtained. The central symbol is the most important in the NLI generation (with red color), and thus, is weighted by $1$. The contribution of the adjacent symbols that are further away by $|l|$ symbol periods, which is determined by $\lambda^{|l|}$, decays exponentially (shown with a faded color). As $i$ changes, the weighted infinite-window will slide through the symbols to obtain a number of weighted energies $G_i^\lambda$.~Then, the EEDI $\hat{\Psi}_\text{Exp}$ is defined as the sample variance over the sample mean of the weighted sum of energy, i.e.,
\begin{equation}\label{expEDI}
\hat{\Psi}_\text{Exp} \define \frac{\TVar{\WE}}{\TExp{\WE}}.
\end{equation}

For $\lambda=1$, since all the symbol energies are weighted by $1$, EEDI is equivalent to EDI with infinite widow, and thus $\hat{\Psi}_\text{Exp}=0$. For $\lambda=0$, $\hat{\Psi}_\text{Exp}$ will converge to $\Exp{|X|^2}(\Phi-1)$, where $\Phi$ represents the standardized fourth moment (a.k.a. kurtosis) of the input symbols. Kurtosis is the NLI predictor proposed by the enhanced GN model, which assumes only i.i.d. transmitted symbols \cite{dar2013properties}. EEDI can be viewed as a refined version of kurtosis that can account for the interaction between non-i.i.d. symbols and the channel memory.

\begin{figure*}[t!]
\centering
\setkeys{Gin}{width=0.24\textwidth}
% \resizebox{0.32\linewidth}{!}{\input{./figures/SNR_EEDIvsN_80km.tikz}}
\includegraphics[width=0.32\linewidth]{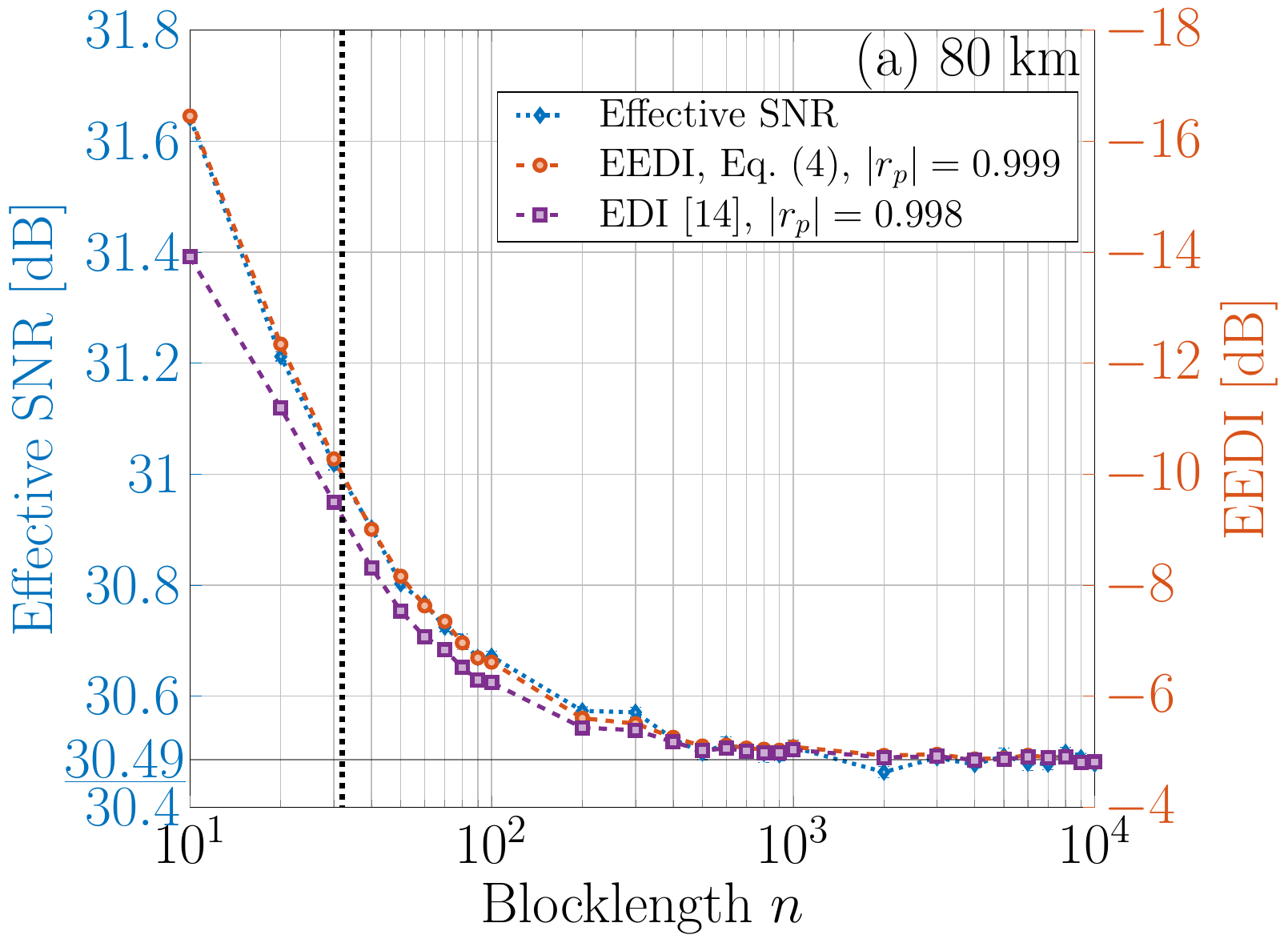}
\hfill
% \resizebox{0.32\linewidth}{!}{\input{./figures/SNR_EEDIvsN_320km.tikz}}
\includegraphics[width=0.32\linewidth]{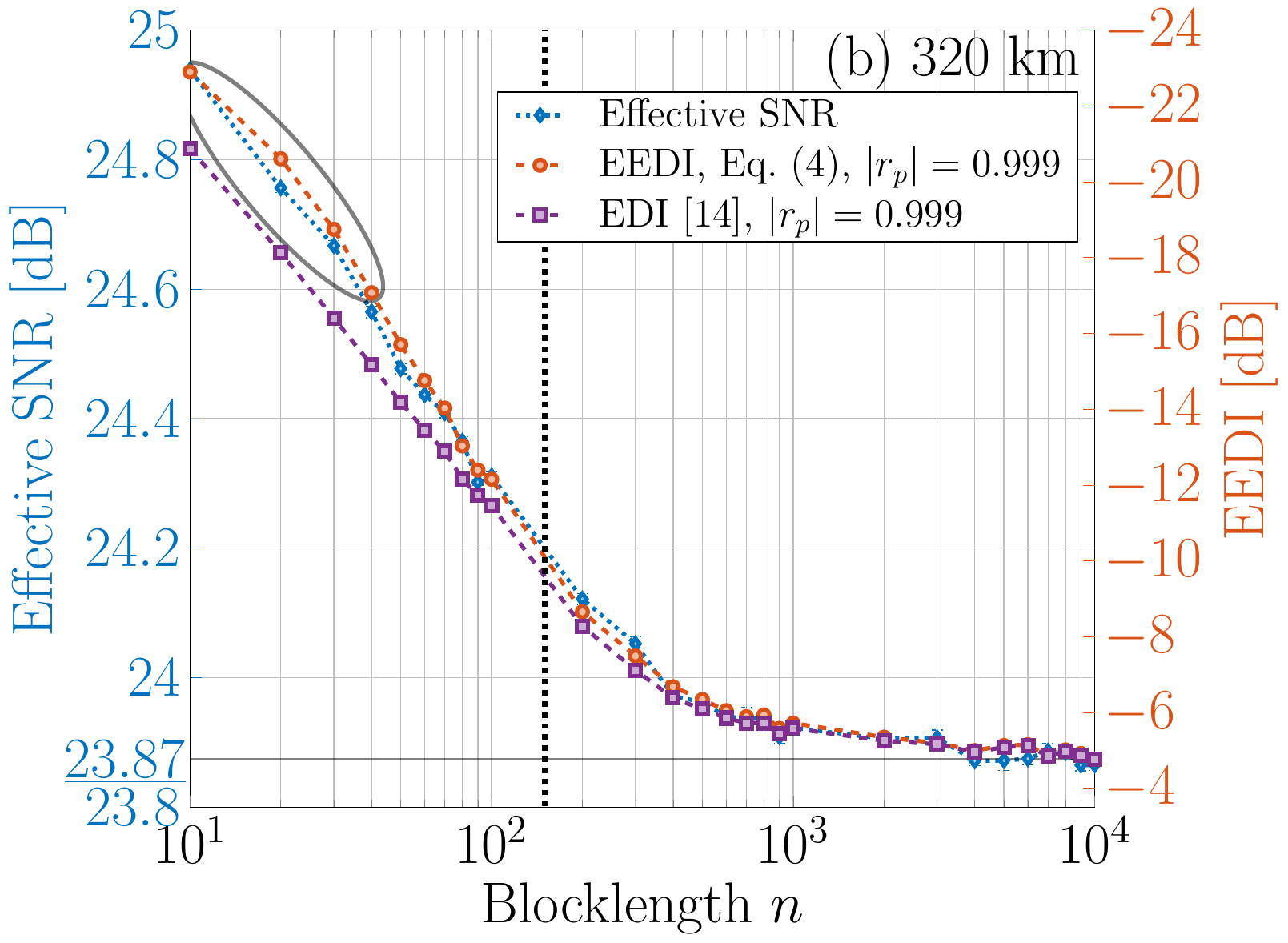}
\hfill
% \resizebox{0.32\linewidth}{!}{\input{./figures/SNR_EEDIvsN_1600km.tikz}}
\includegraphics[width=0.32\linewidth]{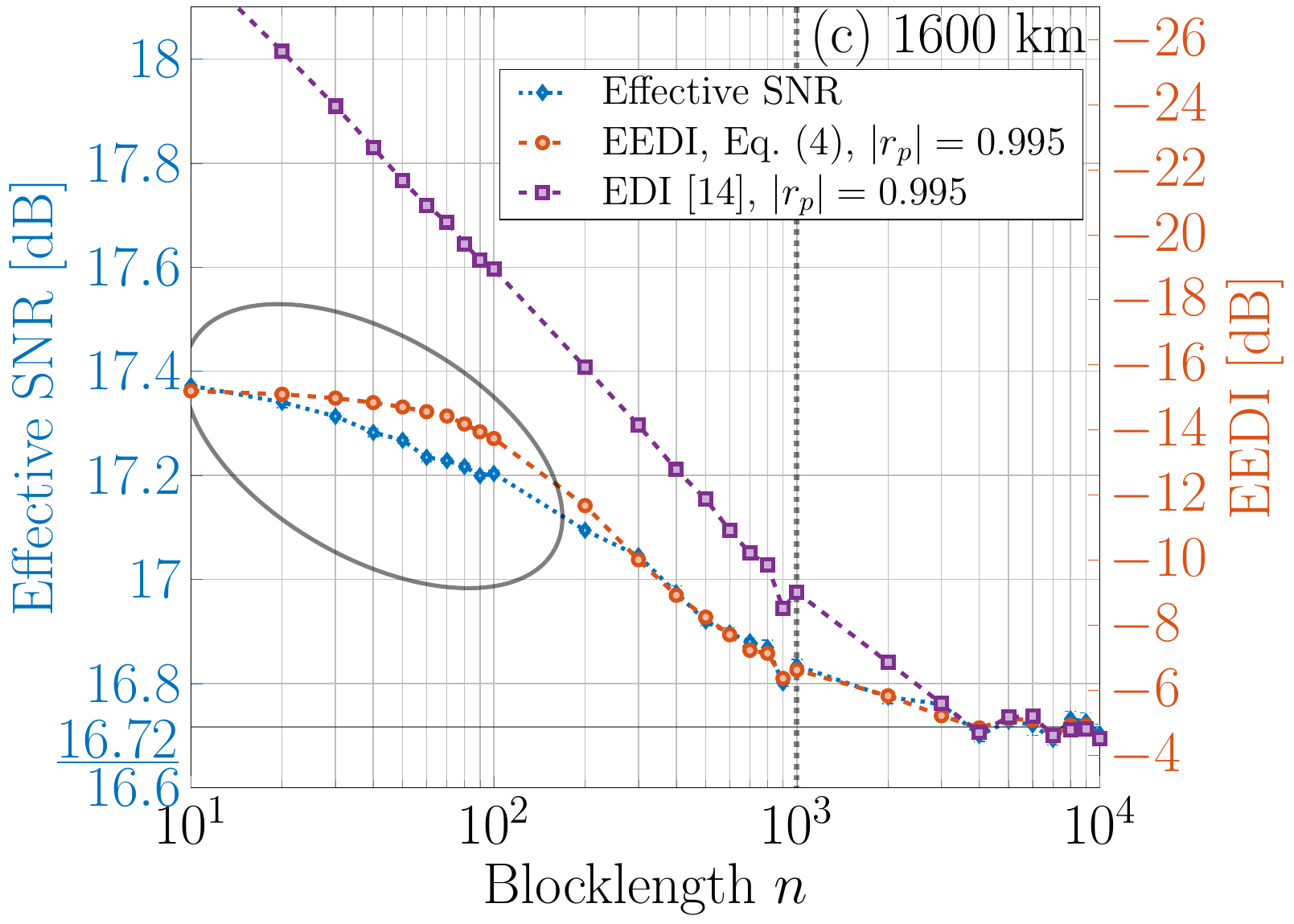}
\caption{Effective SNR (left axis), EEDI, and EDI in \cite{kaiquan2021EDI} (right axis) vs. blocklength. The transmission distances are 80, 320 and 1600 km with the launch powers $-1.5$, $-2.0$ and $-3.0$ dBm, respectively. The EEDI is shown in dB and inverted for convenience of comparison. Error bars for effective SNRs represent 95\% confidence interval. The circled areas in (b)--(c) show the nonlinear behavior of EEDI and effective SNR.}
\label{SNRD}
\end{figure*}

\section{Numerical Results}

Similar to our previous work \cite{kaiquan2021EDI}, we study how well EEDI can predict the blocklength-dependent effective SNR. A single-polarization multi-span wavelength division multiplexing (WDM) system is simulated by using the split-step Fourier method. The system has a span length of 80 km, a fiber loss of 0.19 dB/km, a dispersion parameter of 17 ps/nm/km, and a nonlinear parameter of 1.37 1/W/km. Moreover, a $5\times32$ Gbaud 64-QAM transmission with root-raised-cosine pulse with 10\% roll-off and 50 GHz spacing is considered. The central channel is the one of interest for our analysis. The attenuation after each span is compensated by an erbium-doped fiber amplifier (EDFA) with a noise figure of 6 dB. At the receiver, the channel of interest is processed using ideal chromatic dispersion compensation, matched filtering and sampling. 

For the PAS with 64-QAM, we employ CCDM ranging from ultra short ($n=10$) to long ($n=10,000$) blocklengths.~The amplitude distribution $[0.4,0.3,0.2,0.1]$ is used over the amplitudes $\{1, 3, 5, 7\}$.~The amplitudes on in-phase and quadrature dimensions are independently generated. For each transmission, the same blocklength $n$ is used for all WDM channels.~At the transmitter, we measure the EEDI of the symbols. At the receiver, we evaluate the effective SNR. Pairs of EEDI and effective SNR at different blocklengths are used to obtain their Pearson's correlation coefficient \cite[Ch.~11.1]{gibbons2014nonparametric} $r_p$. The absolute value of coefficient $|r_p|=1$ indicates perfect correlation, while $|r_p|=0$ indicates no correlation.  % between EEDI and effective SNR.

\begin{figure}[!t]
  \centering
        % \resizebox{1\linewidth}{!}{\input{./figures/optFgtFct2.tikz}}
        \includegraphics[width=1\linewidth]{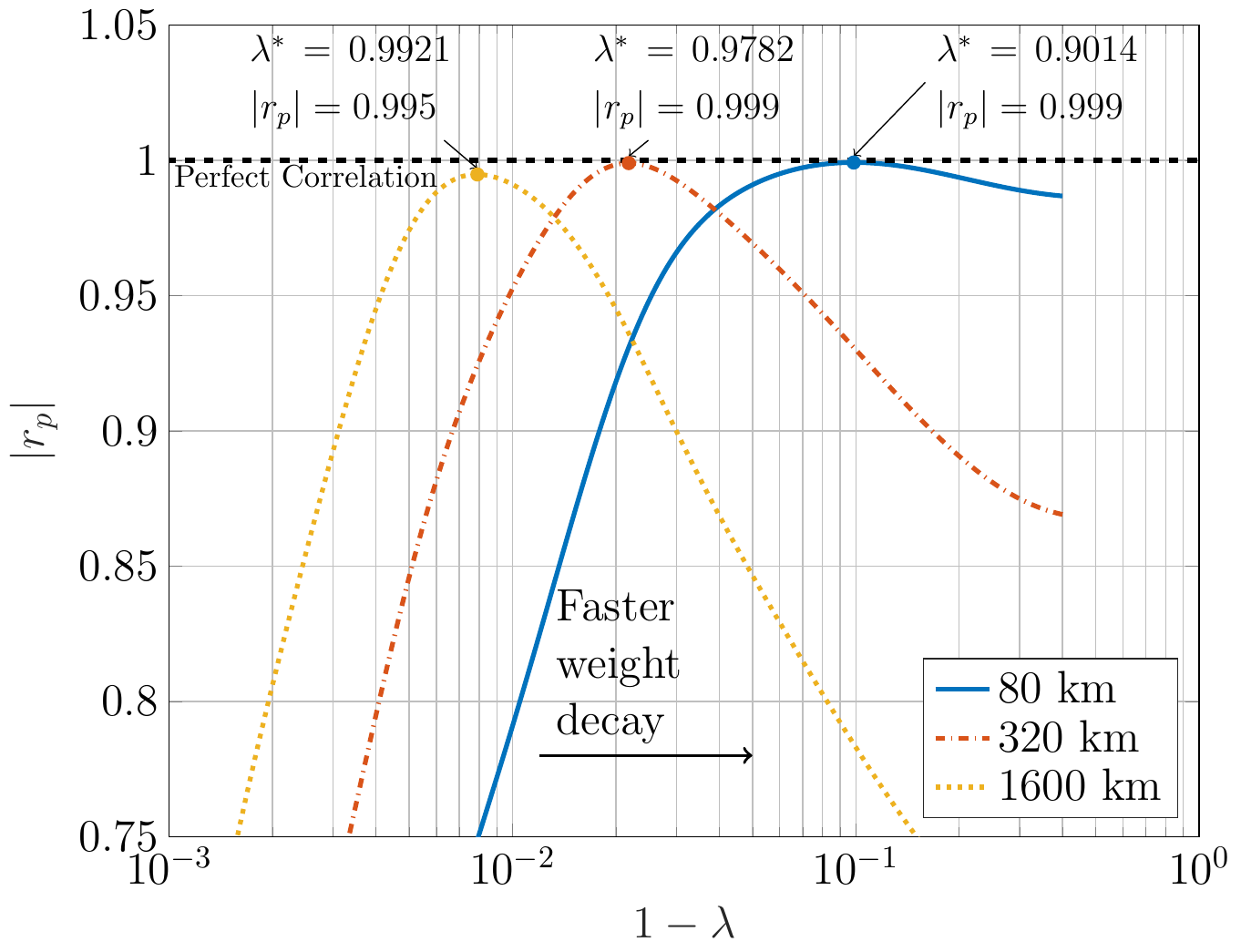}
    \caption{Absolute value of Pearson's linear correlation coefficient $|r_p|$ between EEDI and effective SNR vs. $1-\lambda$. The optimal value of $\lambda$ is denoted by $\lambda^{*}$.}
    \label{optFgtFig}
\end{figure}

EEDI in Fig.~\ref{SNRD} is computed using the optimal forgetting factor, which will be discussed in Fig.~\ref{optFgtFig}. Fig.~\ref{SNRD} shows that both EEDI and EDI predict effective SNR well with absolute correlation coefficient $|r_p|>0.99$ for three distances. To the left of the vertical dotted line, the effective SNR decreases significantly as blocklength $n$ increases. We call this the blocklength-dependent region \cite{kaiquan2021EDI}. Then, the effective SNR decreases slowly until it reaches a floor. For the sake of comparison with the effective SNR, the EEDI is shown in dB and its $y$-axis is inverted, and the EDI from our previous work \cite{kaiquan2021EDI} is shifted vertically by a constant such that it is aligned with EEDI at $n=10,000$. EEDI and EDI have very similar performance in terms of $|r_p|$. However, compared to EDI, one noticeable improvement of EEDI is that the nonlinear decrease of effective SNR at short blocklengths is much better predicted. This nonlinear decrease can be seen by the circled areas in Fig.~\ref{SNRD}~(b)--(c). By contrast, EDI only predicts a linear decay of the effective SNR in these regions. 

The optimal forgetting factor $\lambda^{*}$ used in Fig.~\ref{SNRD} was chosen such that the $|r_p|$ between EEDI and the effective SNR is maximized.~To this end, as shown in Fig.~\ref{optFgtFig}, $\lambda^{*}$ is obtained by exhaustive search from $0.6\leq \lambda < 1$ at a step size of $10^{-4}$.~Note that the $x$-axis in Fig.~\ref{optFgtFig} represents $1-\lambda$, and $|r_p|$ reaches its peak at $1-\lambda^{*}$.~As distance increases from 80 to 1600 km, $\lambda^{*}$ increase from $0.9014$ to $0.9921$, which means that the decaying becomes slower, and more symbols are heavily involved in the nonlinear interaction. For all investigated cases, $|r_p|$ peaks at values very close to $1$, indicating almost perfect correlation between EEDI and effective SNR for $\lambda=\lambda^{*}$.
%that the EEDI becomes slightly less accurate. This could be due to the fact that exponential decaying assumption might not fully describe the decaying behavior of the kernels $S_{h, l, l}$. 
%In general, Fig.~\ref{optFgtFig} shows that EEDI and effective SNR are almost perfectly correlated with each other for $\lambda=\lambda^{*}$.

\begin{figure}[!t]
  \centering
        % \resizebox{1\linewidth}{!}{\input{./figures/optFgtvsDst.tikz}}
        \includegraphics[width=1\linewidth]{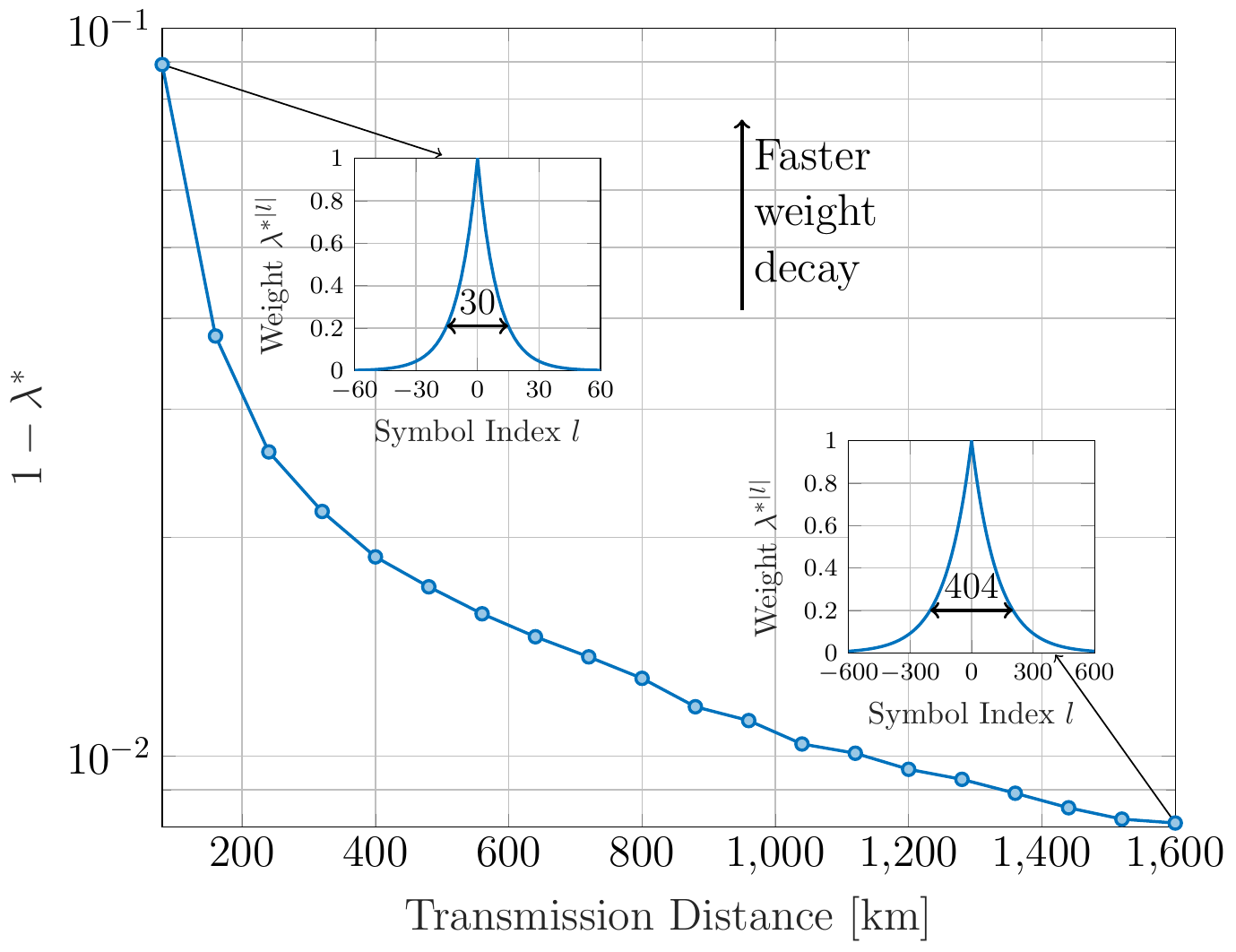}
    \caption{$1-\lambda^{*}$ (all $\lambda^{*}$ obtained with $|r_p|> 0.994$) at distances from 80 km to 1600 km.}
    \label{FgtDstFig}
\end{figure}

Finally, Fig.~\ref{FgtDstFig} shows $1-\lambda^{*}$ at various distances. The $\lambda^{*}$ at each distance is obtained with $|r_p|$ at least $0.994$. Fig.~\ref{FgtDstFig} shows that as the transmission distance increases from 80 km to 400 km, $1-\lambda^{*}$ decreases significantly and begins to decrease at a slower rate. The inset figures of Fig.~\ref{FgtDstFig} show that at 80 and 1600 km, around $30$ and $404$ symbol energies are weighted more than $20\%$, respectively. 

%-------------------------------------------------- Conclusions Section ———————————————————————————%

\section{Conclusions}
We conclude that by using exponential weighting method, EEDI evaluated with the optimal forgetting factor is capable of reflecting the impact of blocklength and distance on the NLI. In addition, EEDI shows superior performance over EDI in terms of predicting the effective SNR for long-distance transmission. Future work will focus on the robustness of the EEDI at longer transmission distances and larger WDM bandwidth.

\section*{Acknowledgments}
The work of K.~Wu, Y.C. G\"{u}ltekin and A.~Alvarado has received funding from the European Research Council (ERC) under the European Union's Horizon 2020 research and innovation programme (grant agreement No 757791). The work of G.~Liga has received funding from the EuroTechPostdoc programme under the European Union's Horizon 2020 research and innovation programme (Marie Sk\l{}odowska grant agreement No 754462).

\ifCLASSOPTIONcaptionsoff
  \newpage
\fi

% trigger a \newpage just before the given reference
% number - used to balance the columns on the last page
% adjust value as needed - may need to be readjusted if
% the document is modified later
%\IEEEtriggeratref{8}
% The "triggered" command can be changed if desired:
%\IEEEtriggercmd{\enlargethispage{-5in}}

% ====== REFERENCE SECTION

%\begin{thebibliography}{1}

% IEEEabrv,

\bibliographystyle{IEEEtran}
\bibliography{references}

\end{document}